\documentclass[aps,prl,twocolumn,superscriptaddress,amsfont,graphicx,nofootinbib,preprintnumbers]{revtex4-1}%
\usepackage{color,graphicx,epsfig}
\usepackage{ifpdf}
\usepackage{amsmath}
\usepackage{bm}
\usepackage[english]{babel}
\usepackage{amssymb}
\usepackage{braket}
\usepackage{hyperref}
\usepackage{enumerate}
\usepackage{url}

\usepackage{slashed}

\usepackage{changes}

\begin{document}

\title{Accelerated Light Dark Matter-Earth Inelastic Scattering in Direct Detection}

\author{Liangliang Su}
\email{liangliangsu@njnu.edu.cn}
\affiliation{Department of Physics and Institute of Theoretical Physics, Nanjing Normal University, Nanjing, 210023, China}

\author{Lei Wu}
\email{leiwu@njnu.edu.cn}
\affiliation{Department of Physics and Institute of Theoretical Physics, Nanjing Normal University, Nanjing, 210023, China}

\author{Ning Zhou}
\email{nzhou@sjtu.edu.cn}
\affiliation{School of Physics and Astronomy, Shanghai Jiao Tong University,
Key Laboratory for Particle Astrophysics and Cosmology (MOE) $\&$ Shanghai
Key Laboratory for Particle Physics and Cosmology, Shanghai 200240, China}

\author{Bin Zhu}
\email{zhubin@mail.nankai.edu.cn}
\affiliation{School of Physics, Yantai University, Yantai 264005, China}

\date{\today}

\begin{abstract}
The Earth-stopping effect plays a crucial role in the direct detection of sub-GeV dark matter. Besides the elastic scattering process, the quasi-elastic and deep inelastic scatterings between dark matter and nucleus that are usually neglected can dominate the interaction, especially in the accelerated dark matter scenarios, which may affect the dark matter detection sensitivity significantly for the underground experiments. We calculate such inelastic scattering contributions in the Earth-stopping effect and illustrate the essence of our argument with the atmospheric dark matter. With the available data, we find that the resulting upper limits on the atmospheric dark matter-nucleus scattering cross-section can differ from those only considering the elastic scattering process by one order of magnitude.
\end{abstract}

\maketitle

\section{Introduction}
There is overwhelming evidence for the existence of dark matter (DM), but the fundamental nature of DM remains a mystery. So far, many well-motivated DM candidates have been proposed, such as weakly interacting massive particles (WIMPs)~\cite{Lee:1977ua,Jungman:1995df}, whose masses vary from GeV$/c^2$ to TeV$/c^2$. Up to now, there has been no conclusive evidence for WIMPs yet from underground experiments~\cite{XENON:2019gfn,PandaX-II:2020oim,LZ:2022ufs}. Beyond the WIMPs, sub-GeV DM with mass below GeV$/c^2$ is another popular thermal candidate and is naturally predicted if the DM couples very weakly to the visible sector~\cite{Essig:2011nj,Essig:2017kqs,Schutz:2016tid,Knapen:2017xzo,DAgnolo:2018wcn,Bertone:2018krk}. On the other hand, a large parameter space with the DM masses in the keV$/c^2$ to GeV$/c^2$ range is still unexplored by conventional direct detection experiments. It facilitates the development of new detection mechanisms and target materials (see recent review e.g.~\cite{Battaglieri:2017aum,Kahn:2021ttr} and references therein). 

Among them, sub-GeV DM with significant Lorentz boosting is of particular interest. Such DM particles can be produced via decays of heavier particles or collisions with energetic cosmic rays, including , for instance, boosted DM (BDM)~\cite{Agashe:2014yua, Berger:2014sqa, Agashe:2015xkj}, solar reflection DM~\cite{An:2017ojc,Emken:2021lgc,An:2021qdl}, cosmic ray boosted DM (CRDM)~\cite{Bringmann:2018cvk,Ema:2018bih,Cappiello:2019qsw, Wang:2019jtk, Guo:2020oum, Ge:2020yuf, Xia:2020apm, Bell:2021xff, Feng:2021hyz, Wang:2021nbf,PandaX-II:2021kai}, and atmospheric DM (ADM)~\cite{Alvey:2019zaa, Su:2020zny, Arguelles:2022fqq, Darme:2022bew,Du:2022hms}. The kinetic energy of these accelerated DM particles can reach up to 1 GeV or even higher, which allows sub-GeV DM particles to induce detectable recoil signals in direct detection experiments.

Note that, after being produced, the accelerated sub-GeV DMs need to pass through the Earth medium to reach underground detectors. Due to the inevitable DM-Earth interaction, they will lose energy and thus get attenuated, which is the so-called Earth-stopping effect~\cite{Kouvaris:2014lpa, Kavanagh:2016pyr, Kavanagh:2017cru}. In previous works, only elastic scattering (ES) is considered in the DM-Earth interaction. This assumption is reasonable for DMs with low kinetic energy. However, for accelerated DMs, the inelastic interactions, including quasi-elastic scattering (QES) and deep inelastic scattering (DIS), can dominate the DM-Earth interaction, which breaks down the ES-only assumption. This problem has been noticed~\cite{Agashe:2014yua,deNiverville:2016rqh,Bringmann:2018cvk,Alvey:2022pad,Kolesova:2022kvq} but is still open. 

With the impulse approximation scheme and the parton model, for the first time, we calculate the QES and DIS contributions in the Earth stopping effect
for accelerated sub-GeV DMs. The sub-GeV ADM model with a scalar mediator is considered, where the collisions of cosmic rays with the atmosphere produce energetic mesons and the mesons then decay into DMs. Such a DM particle obtains a large Lorentz boost from the decay of mesons. To model the DM propagation in the Earth, we take two benchmark models, ``single scattering''~\cite{Kavanagh:2016pyr} and ``straight lines''~\cite{Kavanagh:2017cru, Bringmann:2018cvk}. Including the contributions of the inelastic scattering, we find the new upper bound of the ADM-nucleus scattering cross section can be changed by about one order of magnitude in comparison with that based on the elastic scattering only. Although we focus on the scalar mediator, our argument is general and can be extended to, for instance, vector mediator.

\section{DM-nucleus Inelastic Scattering}

We calculate the DM-nucleus elastic and inelastic scattering in a simplified hadrophilic DM model with a scalar mediator~\cite{Batell:2018fqo}. The relevant DM-quark interactions are given by 
\begin{equation}
\mathcal{L}_I  = g_{\chi}S \bar{\chi}_L\chi_R + g_{u} S \bar{u}_L u_{R} 
\end{equation}
where $g_{\chi}$  and $g_{u}$ are the couplings of mediator $S$ with dark matter and up-quarks, respectively. The corresponding effective Lagrangian of DM-nucleus interaction can be written as~\cite{AristizabalSierra:2018eqm, Batell:2018fqo}
\begin{equation}
\mathcal{L}_I  = g_{\chi}S \bar{\chi}_L\chi_R + g_{A} S \bar{A}_L A_{R} F(Q^2) 
\end{equation}
where $g_{A} = Zg_{pS}+(A-Z)g_{nS}$ is the couplings of mediator $S$ with the nucleus $A$, where $g_{pS} = 0.014 g_u m_p/m_u $ and  $g_{nS} = 0.012 g_u m_n/m_u $ are the couplings of mediator $S$ with proton and neutron, respectively. For simplicity, we assume the isospin to be conservative, $g_{nS} =g_{pS}$. The nuclear form factor, $F(Q^2)$, takes the Helm form factor~\cite{Duda:2006uk} in this work. Then, the differential cross section of DM-nucleus elastic scattering is given by
\begin{equation}
\begin{aligned}
\frac{\mathrm{d} \sigma_{\mathrm{ES}}}{\mathrm{d} E_R} 
& = \frac{ \bar{\sigma}_{\mathrm{n}} A^2 m_S^4F^2(E_R) }{32 \mu_{\mathrm{n}}^2 m_A(2m_A E_R+m_S^2)^2 (E_{\chi}^2-m_{\chi}^2)}\\ 
&\times (4 m_{\chi}^2+2m_A E_R)(4 m_{A}^2+2m_A E_R),
\end{aligned}
\label{eq:elastic_xs}
\end{equation}
where $E_{\chi}$ is the incoming DM energy and $\mu_{\mathrm n}$ is the reduced mass of DM and nucleon. To compare with the experiment data, we define a momentum-independent DM-nucleon scattering cross section $\bar{\sigma}_{\mathrm{n}}={g_{\chi}^2g_{pS}^2 \mu_{\mathrm n}^2}/{\pi m_S^4}$. The recoil energy $E_R = Q^2 /2m_A$ is the function of momentum transfer $Q$ and nucleus mass $m_A$. Such an assumption is reasonable as the inverse of momentum transfer $|\vec{q}|$ to the scatterer is larger than the radius of the scatterer. However, for a sub-GeV DM with a large boost, the QES and DIS processes must be considered in the high kinetic energy region. In the former, one or more nucleons are dislodged or excited inside atom $A$, but in the latter, the nucleus will disintegrate into a large number of hadrons.

{\it Deep Inelastic Scattering:} Under the parton model, the DM-nucleus DIS can be simplified to $\chi (k) +q(x p) \to \chi({k^{\prime}}) +q^{\prime} (p^{\prime}) $, where $x = Q^2/(2 m_A \nu)$ is defined as the Bjorken scaling variable. It is a function of transfer energy $\nu$ and the square of transfer momentum $Q^2 \equiv-q^2 = -(k-k^{\prime})^2=2E_{\chi}(E_{\chi}-\nu)-2|\vec{k}||\vec{k}|^{\prime}\cos{\theta}-2m_{\chi}^2$, where $\theta$ is the scattering angle between DM and quarks. In the rest frame of the target particle, the differential cross section of DIS is given by 
\begin{equation}
\begin{aligned}
\mathrm{d} \sigma_{\mathrm{DIS}} 
&=\frac{\mathrm{d}\nu\mathrm{d}Q^2}{64 \pi m_A^2\nu (E_{\chi}^2-m_{\chi}^2)}\int_{0}^{1}\frac{f(\xi)}{\xi}\mathrm{d} \xi \overline{|\mathcal{M}(\xi)|^2}\delta(\xi-x)\\
& = \sum_{q} \frac{g_{\chi}^2g_q^2(4 m_{\chi}^2 + Q^2)(4m_q^2+ Q^2)\mathrm{d}\nu\mathrm{d}Q^2}{32 \pi  m_A  Q^2(E_{\chi}^2-m_{\chi}^2)(Q^2+m_S^2)^2}f_{q/A}(x,Q^2),\\
\end{aligned}
\end{equation}
where $q = \{u,\bar{u}\}$ in our calculations. $\overline{|\mathcal{M}(\xi)|^2}$ is the square of spin-averaged amplitude of DM-quarks scattering. The function $f_{q/A}(x,Q^2)$ is the nuclear parton distributions (nPDFs)~\cite{Buckley:2014ana,AbdulKhalek:2022fyi}.

{\it Quasi-elastic Scattering:} at moderate incident energy, the DM elastically scatters with the quasi-free nucleons bounded in a nucleus, 
\begin{equation}
    \chi (k) + A(p_A) \to \chi(k^{\prime}) + X(\to N+ Y). 
\end{equation}
Here $N$ and $Y=(A-1)$ denote nucleon and residual nucleus, respectively. In Born approximation, the double differential cross section of DM-nucleus QES via a scalar mediator can be given by
\begin{equation}
\frac{\mathrm{d} \sigma_{\mathrm{QE}}}{\mathrm{d} E_{\chi}^{\prime} \mathrm{d} \Omega}=\frac{\bar{\sigma}_{\mathrm{n}} m_S^4}{16 \pi \mu_{\mathrm{n}}^2} \frac{\left|\vec{k}^{\prime}\right|}{|\vec{k}|} \frac{\mathcal{X}_{S} W_{S}}{(Q^2+m_S^2)^2},
\end{equation}
where $E_{\chi}^{\prime}$ is the outgoing DM energy. The DM tensor and nuclear tensor, $\mathcal{X}^S$ and $W^S$, are defined as 
\begin{equation}
\begin{aligned}
\mathcal{X}^{S} =&\overline{\sum}\left\langle \chi\left|j^{S}_{\chi}\right| \chi^{\prime}\right\rangle\left\langle \chi^{\prime}\left|j^{S}_{\chi}\right| \chi \right\rangle =  4 m_{\chi}^2+Q^2; \\
W^{S} =&\overline{\sum}\left\langle A\left|J^{S}(0)\right| X\right\rangle\left\langle X\left|J^{S}(0)\right| A\right\rangle \\
&\times\delta^{(4)}\left(p_X+k^{\prime}-p_A-k\right),
\end{aligned}
\end{equation}
where $j_{\chi}^{S}$ and $J^{S}$ are the DM and nuclear scalar currents operator, respectively. The nuclear tensor $W^{S}$ includes all the information about the structure of the target nucleus. In the low momentum transfers, $W^{S}$ can be obtained by the nuclear many-body theory (NMBT)~\cite{Carlson:1997qn} which regards the initial and final states as non-relativistic wave functions, and the current operator is expanded by the Taylor series of $|\vec{q}|/m_N$. However, the non-relativistic wave functions are improper for the final state $|X\rangle$ with high momentum transfer, for instance, the incident energy of DM is larger than several hundred MeV.

The impulse approximation (IA) is an excellent scheme to calculate the inclusive cross-section of QES for the high momentum transfer~\cite{Benhar:2005dj, Ankowski:2005wi, Ankowski:2007uy, Ankowski:2011ei, Ankowski:2013gha}. It assumes that {\it i}) the DM-nucleus scattering is reduced to the incoherent sum of the scattering processes involving individual nucleons; {\it ii}) the nucleon $N$ and residual nucleus $(A-1)$ after scattering are independent. We neglect the dynamical final state interactions (FSI), but consider the effect of Pauli blocking in this work.

\begin{figure}[ht]
  \centering
  \includegraphics[width=7cm,height=5cm]{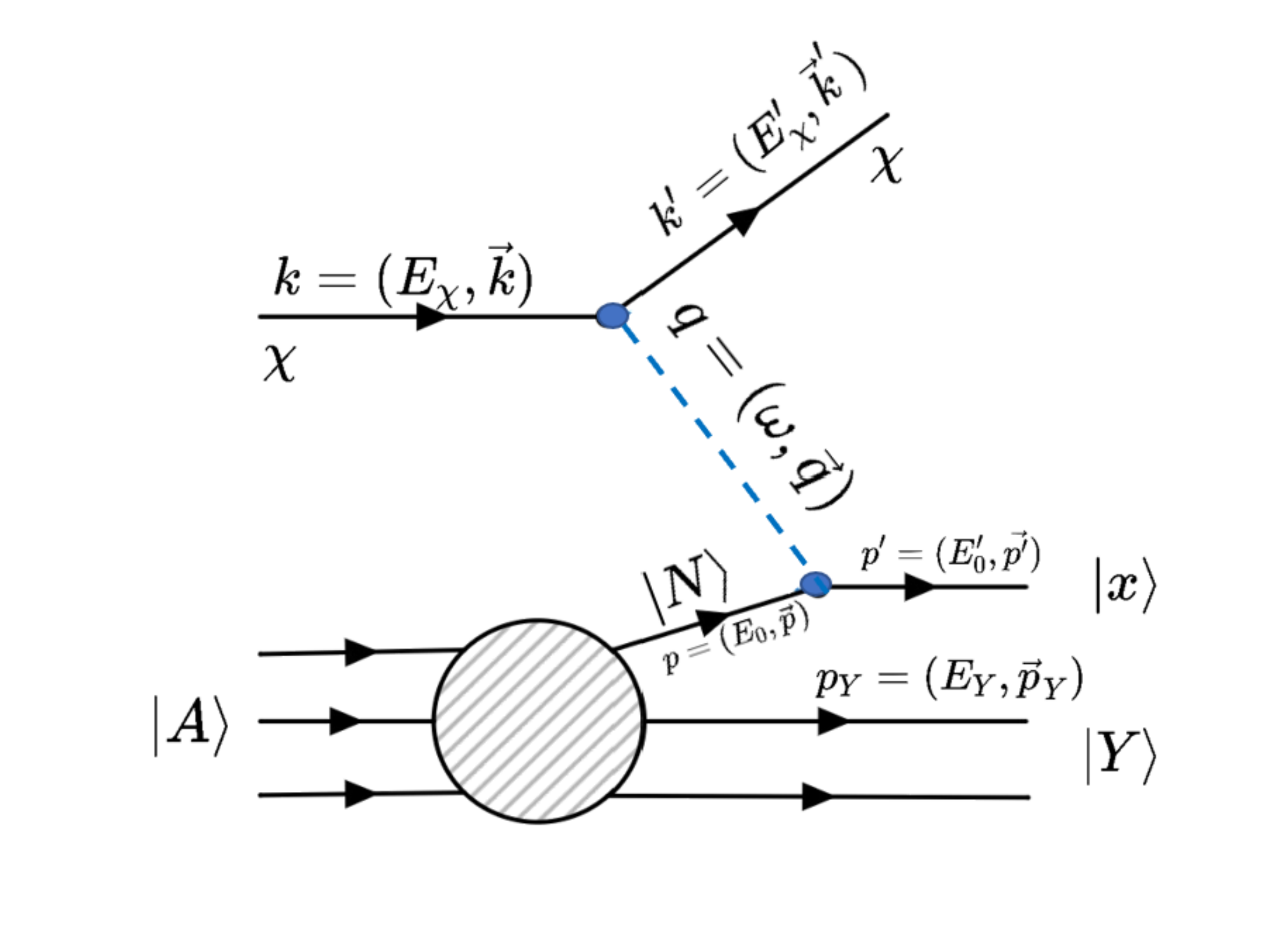}
  \caption{The diagrammatic sketch of DM-nucleus quasi-elastic scattering under the IA scheme.}
  \label{fig:IA}
\end{figure}
Under the IA scheme, the nuclear current operator $J^{S}(0)$ can be calculated as the sum of individual nucleon currents  $J^{S} \to \sum_{N} j_{N}^{S} $, and the final state $|X\rangle$ can be separated to the knockout nucleon $|x,\vec{p}^{\prime}\rangle$ and the residual nucleus $|Y,\vec{p}_Y\rangle$, as shown in Figure~\ref{fig:IA},
\begin{equation}
|X\rangle \rightarrow|x, \vec{p}^{\prime}\rangle \otimes\left|Y, \vec{p}_{Y}\right\rangle.
\end{equation}

\begin{figure*}[ht]
  \centering
\includegraphics[height=6cm,width=17.5cm]{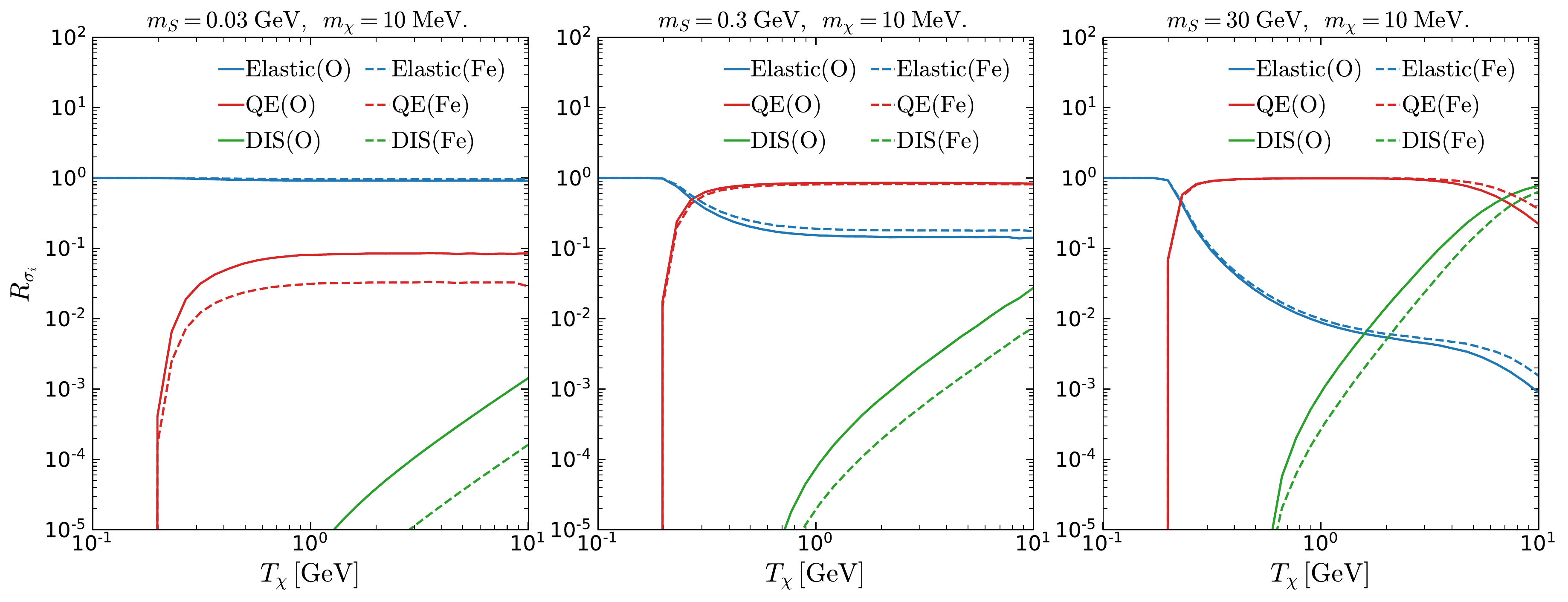}
\caption{The ratio $R_{\sigma_i}$ (c.f. Eq.~\ref{eq:ratio}) as the function of DM kinetic energy $T_\chi$ at the DM mass $m_\chi=10$ MeV for different mediator masses $m_S=0.03$ GeV (left panel), 0.3 GeV (middle panel) and 30 GeV (right panel). The solid and dotted lines denote the DM-Oxygen and DM-Iron scattering, respectively.}
\label{fig:sigma_tot}
\end{figure*}

The inclusive differential cross-section of the DM-nucleus QES can be given by
\begin{equation}
 \frac{\mathrm{d} \sigma_ {\mathrm{QE}}}{\mathrm{d} E_{\chi}^{\prime} \mathrm{d} \Omega}   = Z\frac{\mathrm{d} \sigma_p}{\mathrm{d} E_{\chi}^{\prime} \mathrm{d} \Omega}+(A-Z)\frac{\mathrm{d} \sigma_n}{\mathrm{d} E_{\chi}^{\prime} \mathrm{d} \Omega},
\end{equation}
with 
\begin{equation}
\begin{aligned}
  \frac{\mathrm{d} \sigma_N}{\mathrm{d} E_{\chi}^{\prime} \mathrm{d}  \Omega} &= \frac{\bar{\sigma}_{\mathrm{n}} m_S^4}{16 \pi\mu_{\mathrm{n}}^2(Q^2+m_S^2)^2} \frac{\left|\vec{k}^{\prime}\right|}{|\vec{k}|} \int \mathrm{d}^{3} \vec{p} \mathrm{~d} E \frac{m_N^2}{E_{\vec{p}}E_{ \vec{p}^{\prime}}}  P(\vec{p}, E)\\ &\times \Theta(|\vec{p}^{\prime}|-p_F) \delta\left(\omega-E+m_{N}-E_{0}^{\prime}\right)  \mathcal{X}^{S} {H}^{S}_N.
\end{aligned}
\label{eq:d_QE}
\end{equation}
Here $N= \{p,n\}$ and $\Theta(|\vec{p}^{\prime}|- p_F)$ come from the nuclear Pauli blocking, and $p_F$ is Fermi momentum. $m_N/E_{\vec{p}}$ and  $m_N/E_{\vec{p}^{\prime}}$ are the covariant normalization factors. 
The hadronic tensor, ${H}^{S}$, is defined by,
\begin{equation}
\begin{aligned}
 {H}^{S}_N &=\overline{\sum}\left\langle N,-\vec{p}\left|j_{N}^{S}\right| x, \vec{p}+\vec{q}\right\rangle\left\langle\vec{p}+\vec{q}, x\left|j_{N}^{S}\right| N, \vec{p}\right\rangle \\
 & = \frac{1}{2}\mathrm{Tr}\left[\Gamma^S\frac{\not p+m_N}{2m_N} \Gamma^{S\dagger} \frac{\not p^{\prime}+m_N}{2m_N} \right]
\end{aligned}
\end{equation}
with
\begin{equation}
\Gamma^{S} = F_S(Q^2) = \frac{\xi_S}{(1 +Q^2/\Lambda^2_S)^2},
\end{equation}
where $F_S(Q^2)$ is the scalar nucleon form factor~\cite{formfactor}. We take $\xi_S =1.8$ and $\Lambda_S = 1.0 \;\mathrm{GeV} $~\cite{Kuzmin:2004ke}. It should be noted that the transfer momentum $Q$ in Eq.~\ref{eq:d_QE} is not only transferred to the interacting nucleon, but also the residual nucleus system. Thus, we have to handle the problem with the off-shell kinematics~\cite{DeForest:1983ahx}, i.e., $q \equiv (\omega, \vec{q}) \to \tilde{q} \equiv (\tilde{\omega},\vec{q})$, where $\tilde{\omega} = E_{\vec{p}^{\prime}}-E_{\vec{p}}= \omega -E+m_N-E_{\vec{p}}$ and $\omega$ is the transfer energy. The spectral function of the target nucleus, $P(\vec{p}, E)$ in Eq.~\ref{eq:d_QE}, represents the probability of removing a nucleon with momentum $\vec{p}$ and removal energy $E$ from the bound state of the nucleus~\cite{Benhar:1994hw, Benhar:2005dj},
\begin{equation}
\begin{aligned}
P(\vec{p}, E) &=\sum_{Y}|\langle A \mid Y,-\vec{p}\rangle| N, \vec{p}\rangle|^{2} \\
& \times \delta\left(E-m_{N}+E_{0}-E_{Y}\right).
\end{aligned}
\end{equation}

To compare the contributions of the elastic and inelastic scattering processes, we define the ratio of scattering cross sections, $R_{\sigma_i}$,
\begin{equation}
R_{\sigma_i}=\frac{
\sigma_i
}{\sigma_{tot}},
\label{eq:ratio}
\end{equation}
where $i={\rm ES, QES, DIS}$. In Figure~\ref{fig:sigma_tot}, we show the ratio $R_{\sigma_i}$ as the function of the DM kinetic energy $T_\chi$ for different mediator masses. We consider the oxygen (solid lines) and iron (dotted lines) nuclei in the Earth. It can be seen that the contribution of each process depends on the scalar mediator mass. For instance, when $m_S = 0.03$ GeV, the cross section of the elastic scattering (blue lines) is always larger than the inelastic scattering (red and green lines). However, for $m_S = 0.3$ GeV$/c^2$, the QES becomes the dominant contribution in $ T_{\chi} \gtrsim 200$ MeV range. If $m_S=30$ GeV$/c^2$, the DIS is non-negligible when $T_{\chi} \gtrsim 1$ GeV. These results can be understood as follows: the elastic and inelastic scattering processes occur in the low and high momentum transfer $Q$ regions, respectively. When the mediator mass $m_S$ is much smaller than the typical value of $Q_{\rm ES}$, the elastic scattering cross section can be enhanced by $\sim 1/Q^4_{\rm ES}$ as compared with the inelastic scattering processes, due to $Q_{\rm ES} \ll Q_{\rm QES, DIS}$. On the other hand, if the mediator mass $m_S$ is much larger than $Q_{\rm ES}$, this enhancement in elastic scattering disappears. Such behaviors also appear in the DM-nucleus scattering via exchanging a dark photon, where the QES and DIS are dominant when the dark photon mass is greater than $\mathcal{O}(1)$ GeV.

\section{Earth Stopping}

As the DM particles travel through the Earth toward to the detector, they can interact with the different nuclei species in the Earth, which makes them slow down, or even stop. Such an effect is significant for fast-moving light DM. In our study, we take the ADM as a benchmark model, in which the DM has a large Lorentz boost and thus encounters sizable inelastic scattering with the nucleus in the Earth.

The ADM is produced by the inelastic collision between the cosmic rays (CRs) and the atmosphere on Earth, i.e., 
\begin{equation}
    p+N\to M \to \chi \bar{\chi} +X.
\end{equation}
In our simulation, we only include the contribution of the proton ($p$) in CRs colliding with the nitrogen ($N$) in the atmosphere. The produced mesons, $M$, from this collision will promptly decay to DM pair $\chi \bar{\chi}$ and other SM particles $X$ via an on-shell scalar mediator $S$. We consider $\eta$ meson decay process, $\eta \to \pi^{0} S(\to \chi 
\bar{\chi})$, which requires the mediator mass $m_S$ to satisfy $2m_{\chi}<m_S < m_{\eta} -m_{\pi^{0}}$. Besides, given the constraints from the MINIBooNE experiment and the kaon meson decays~\cite{MiniBooNEDM:2018cxm,BNL-E949:2009dza}, we adopt the appropriate parameters, $m_S = 300$ MeV$/c^2$ and $\mathrm{Br}[\eta \to \pi S (\to \chi \bar{\chi})] = 10^{-5}$. With this setup, we calculate the differential flux of ADM on the surface of the Earth as Ref.~\cite{Alvey:2019zaa,Su:2020zny}.

Then, we take two benchmark Earth-stopping models to show the effects of inelastic scattering on the flux of the ADM reaching the detector. The conservative one assumes that the ADM scatters with nuclei at most once, i.e., the ``single scatter'' approximation~\cite{Kavanagh:2016pyr}. The Earth's rotation effect can be neglected for the fast-moving DM. Thus, the differential flux of the ADM around the detector is given by
\begin{equation}
\begin{aligned}
\frac{\mathrm{d} \Phi_{\chi}^{z}}{\mathrm{d} T_{\chi}^{z}} &= \int  \mathcal{P}_{\mathrm{surv}}(T_{\chi}, \cos \theta)  \frac{\mathrm{d} \Phi_{\chi}}{\mathrm{d} T_{\chi} \mathrm{d} \Omega } \mathrm{d} \Omega\\
&= 2 \pi \int_{-1}^{1}  \mathcal{P}_{\mathrm{surv}}(T_{\chi}, \cos \theta)  \frac{\mathrm{d} \Phi_{\chi}}{\mathrm{d} T_{\chi}\mathrm{d} \Omega } \mathrm{d} \cos \theta,
\end{aligned}    
\end{equation}
where $T_{\chi}^z$ is the kinetic energy of the ADM at the detector. $\mathcal{P}_{\mathrm{surv}}$ is the survival probability of the ADM as it reaches the detector, which is defined as
\begin{equation}
\begin{aligned}
\mathcal{P}_{\mathrm{surv}}(T_{\chi}, \cos \theta) 
&=\mathrm{exp} \left(-\sum_{i} \frac{d_{\mathrm{eff},i}(\cos \theta)}{\bar{\lambda}_i(T_{\chi})} \right),
\end{aligned}
\end{equation}
where $\theta$ is the angle between DM incoming direction and the Earth's core/detector axis.  $\bar{\lambda}_i = [\sigma_i^{tot} (T_{\chi})\bar{n}_i]^{-1}$ is the average mean free path, and $n_i (\bar{n}_i)$ is (average) number density of Earth species $i$. The effective Earth-crossing distance, $d_{\mathrm{eff},i}(\cos \theta)$, is defined by 
\begin{equation}
d_{\mathrm{eff},i} \approx  \left\{\begin{matrix}
 \displaystyle\int_{R_E \sin\theta}^{R_E}  \frac{2r n_i(r) \mathrm{d} r}{\bar{n}_i\sqrt{r^2-R_E^2\sin^2\theta}} ;& \theta \in \left[0,\dfrac{\pi}{2}\right]\\
 \displaystyle\int_{R_E-z_D}^{R_E} \frac{n_{i}(r)}{\bar{n}_i} \mathrm{d} r ,& \theta \in \left[\dfrac{\pi}{2},\pi\right]
\end{matrix}\right.
\end{equation}
where $R_E = 6378.14 \; \mathrm{km}$ and $z_D = 1.4 \;\mathrm{km}$ are the Earth's radius and the depth of Xenon1T experiment, respectively. 

The other model assumes that the DM particles travel in straight lines and lose energy due to the DM-Earth scattering, which we refer to as ``straight lines'' mode~\cite{Kavanagh:2017cru}. Compared with the ``single scatter'' model, this model gives an optimistic prediction of the ADM differential flux around the detector, which is given by 
\begin{equation}
\frac{\mathrm{d} \Phi_{\chi}^{z}}{\mathrm{d} T_{\chi}^{z}} = \int  \frac{\mathrm{d} T_{\chi}}{\mathrm{d} T_{\chi}^{z}}  \frac{\mathrm{d} \Phi_{\chi}}{\mathrm{d} T_{\chi} \mathrm{d} \Omega } \mathrm{d} \Omega.
\end{equation}
Here $\mathrm{d} T_{\chi}/{\mathrm{d} T_{\chi}^{z}}$ can be obtained by solving the energy loss function~\cite{Bringmann:2018cvk, Alvey:2022pad},
\begin{equation}
\frac{\mathrm{d} T_{\chi}^{z}}{\mathrm{d} z} =  -\sum_{i} n_{i}(r) \int_{0}^{\omega_{\chi}^{\max}} \mathrm{d} \omega_{\chi} \frac{\mathrm{d} \sigma_{ \chi i}}{\mathrm{d} \omega_{\chi}} \omega_{\chi},
\end{equation}
where the energy loss $\omega_{\chi}$ is equal to the nuclear recoil energy $E_R$ in the elastic scattering. The differential cross section of the inelastic scattering can be calculated by $\mathrm{d} \sigma_{\chi i}/\mathrm{d} \omega = \int_{Q^2} \dfrac{ \mathrm{d} \sigma_{\chi i}}{\mathrm{d} \omega \mathrm{d} Q^2} \mathrm{d} Q^2 $. Although there are some more accurate Monte Carlo simulations of DM trajectories~\cite{ Emken:2017qmp,Emken:2018run,Mahdawi:2018euy,Emken:2019hgy,Chen:2021ifo,Xia:2021vbz,CDEX:2021cll}, these two benchmark models are enough to show the effects of inelastic scattering in the Earth-stopping.

\begin{figure}[ht!]
\centering
\includegraphics[width=7cm,height=7cm]{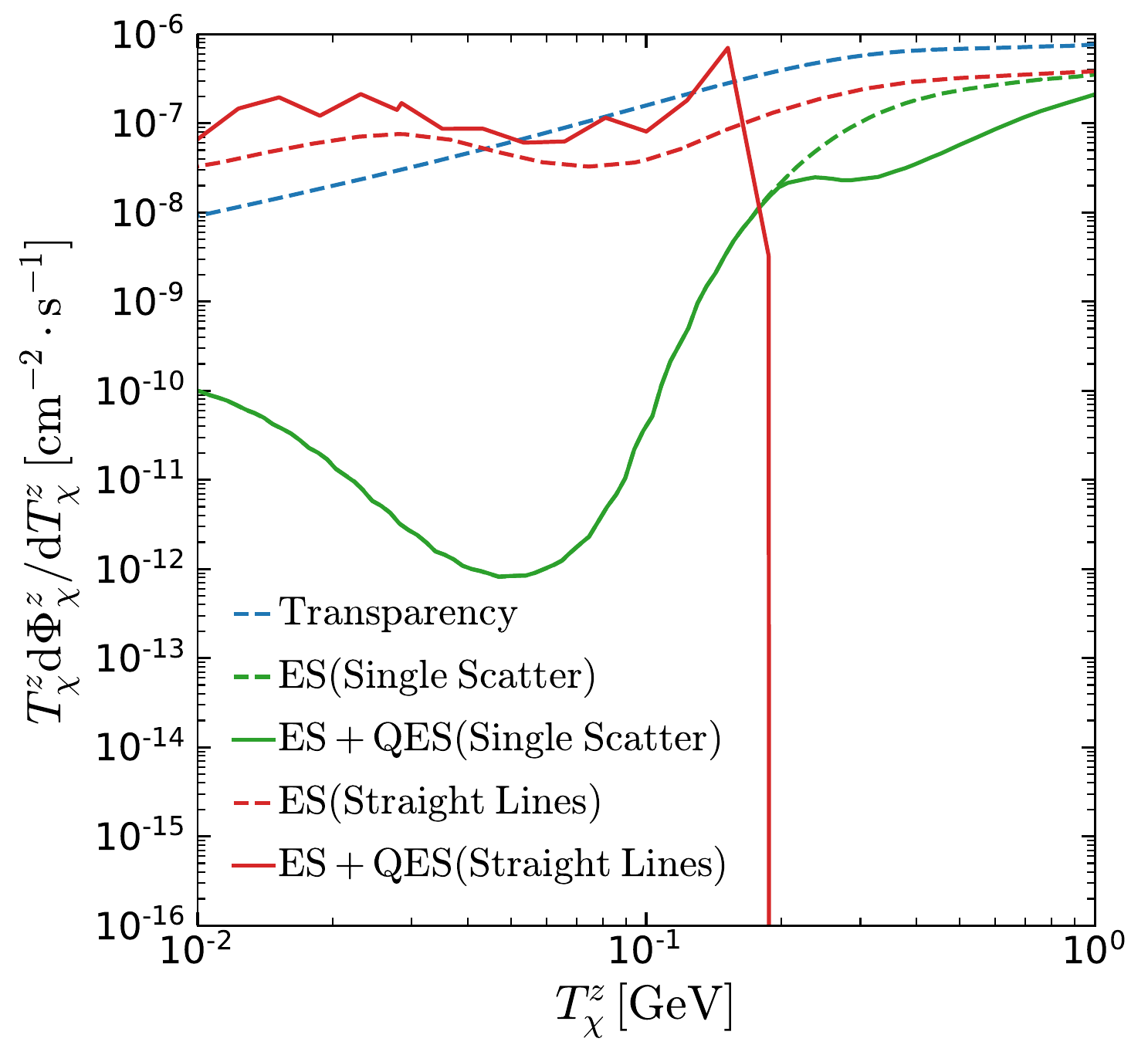}
\caption{The expected differential flux of ADM at the Xenon1T experiment for ``single scatter'' (green lines) and ``straight lines'' (red lines) Earth-stopping models without (dotted lines) and with (solid lines) QES. Here we assume $m_{\chi} =0.01$ GeV$/c^2$, $m_S = 300$ MeV$/c^2$, $\bar{\sigma}_{\mathrm n} = 5 \times 10^{-29} \; \mathrm{cm}^2$, and $\mathrm{Br}[\eta \to \pi S (\to \chi \bar{\chi})] = 10^{-5}$. The result in the transparent Earth case is also plotted (blue dotted line).}
\label{fig:flux}
\end{figure}

Figure~\ref{fig:flux} shows the expected differential flux of ADM reaching the Xenon1T detector for the ``single scatter'' and the ``straight lines'' models with and without the contributions of inelastic scattering. As a comparison, we also present the result under the assumption of transparent Earth. We note that the number of DIS events is negligible in the ADM, and thus focus on the QES. It can be seen that the contribution of QES in both Earth-stopping models becomes sizable in the DM kinetic energy region, $T_{\chi} \gtrsim 200$ MeV, which is consistent with the results in Figure~\ref{fig:sigma_tot}. Including the QES can enhance the DM-Earth scattering cross section, and thus reduces the ADM flux at the detector in the high $T_\chi$ region. On the other hand, in the ``straight lines'' model, a fraction of highly boosted ADM particles that involve in the QES will lose energy and then make the flux of the ADM in low energy region larger than that in the transparent Earth case. Besides, the ADM flux is greatly reduced in $T_{\chi} \in [0.01,0.1]$ GeV because the DM-nucleus ES cross section is enhanced by the momentum transfer effect of the light DM \cite{Flambaum:2020xxo}. While for the larger $T_{\chi}$, the ES cross section is highly suppressed by the nuclear form factor.

\section{Exclusion Limits}
\begin{figure}[ht]
\centering
\includegraphics[width=7cm,height=7cm]{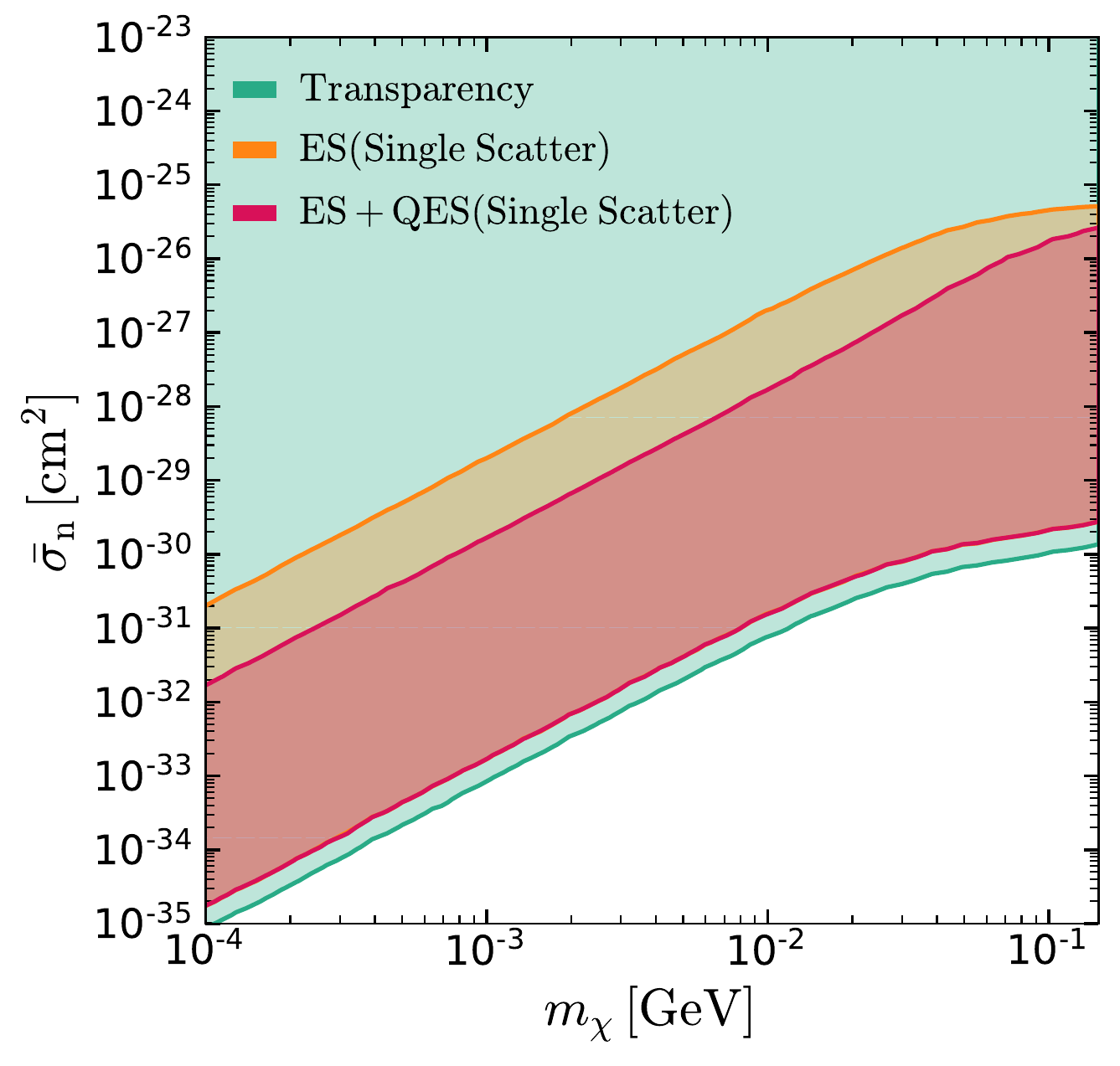}
\includegraphics[width=7cm,height=7cm]{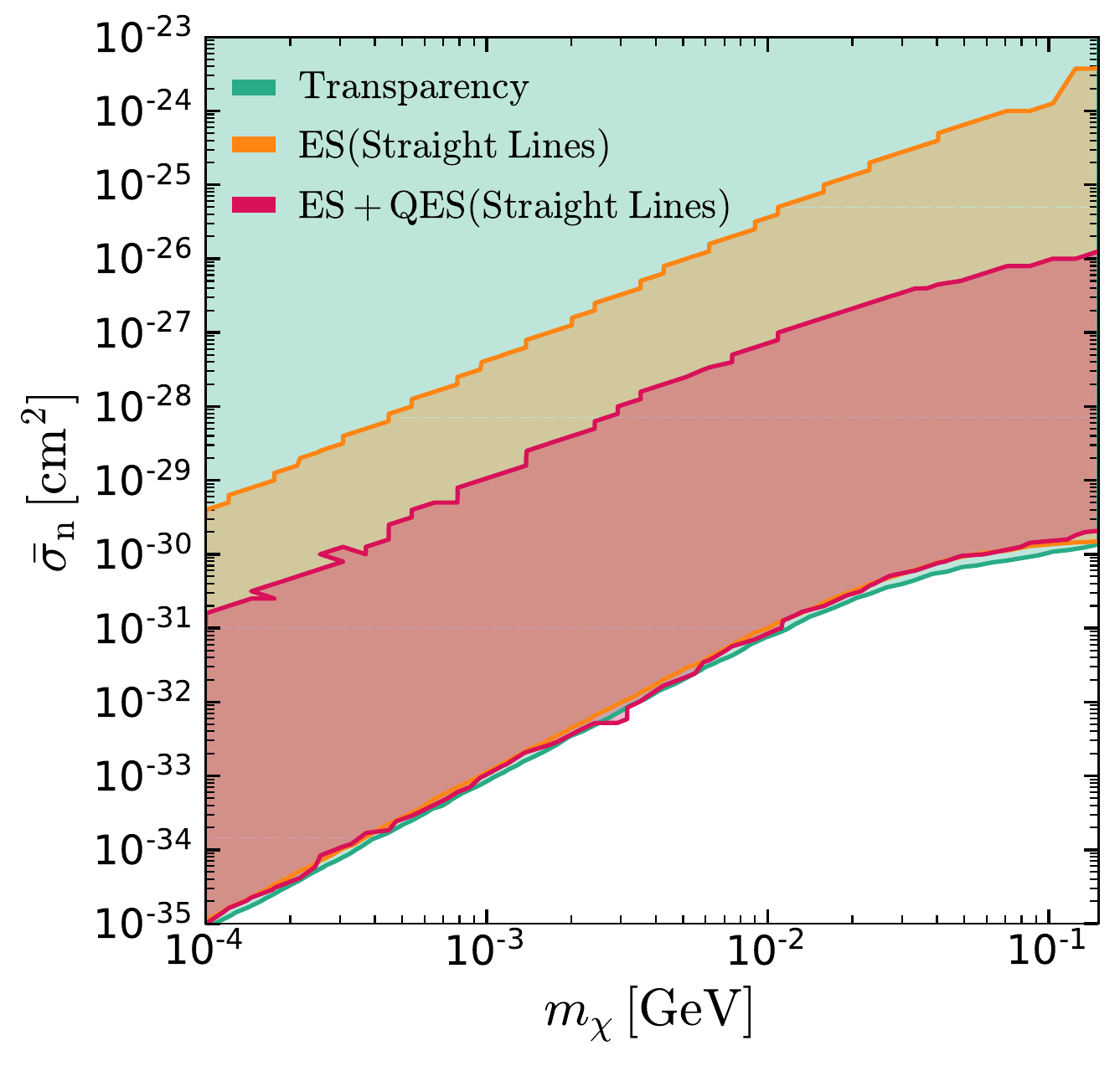}
\caption{The exclusion limits on the momentum-independent ADM-nucleon scattering cross section $\bar{\sigma}_{\mathrm n}$ versus the DM mass $m_\chi$. The green, orange and red region denotes the results for the transparent Earth, Earth-stopping with the elastic scattering only and with elastic scattering plus quasi-elastic scattering, respectively. We assume $m_S = 300 $ MeV and $\mathrm{Br}[\eta \to \pi S (\to \chi \bar{\chi})] = 10^{-5}$. Two Earth-stopping models ``Single Scatter'' (upper panel) and ``Straight Lines'' (lower panel) are considered.} 
\label{fig:sigma_limit}
\end{figure}

With the above differential flux, we can evaluate the nuclear recoil rate of the ADM in Xenon1T experiment,
\begin{equation}
R = N_T \int_{E_R^{\min}}^{E_R^{\max}}  \mathrm{d} E_R \int_{T_{\chi}^{z,\min}}^{T_{\chi}^{z,\max}} \epsilon(E_R) \frac{\mathrm{d} \Phi_{\chi}^{z}}{\mathrm{d} T_{\chi}^{z}} \frac{\mathrm{d} \sigma}{\mathrm{d} E_R} \mathrm{d} T_{\chi}^{z}
\end{equation}
where $N_T$ and $\epsilon$ are the number density of Xenon and the detector efficiency with the nuclear recoil energy $E_R$, respectively. We use Xenon1T data in the energy range, 4.9 keV $< E_R<$ 40.9 keV~\cite{XENON:2018voc}, to derive the exclusion limits. In Figure~\ref{fig:sigma_limit}, we show the exclusion limits on the momentum-independent ADM-nucleon scattering cross section $\bar{\sigma}_{\mathrm n}$ in the cases of the transparent Earth, Earth-stopping with the ES only and with ES plus QES. For both models, we can see that there are upper bounds on scattering cross section $\bar{\sigma}_n$ because of the Earth-stopping effect. Besides, comparing with the ES only, we find that the upper bounds including the DM-Earth QES can be changed by about one order of magnitude. On the other hand, the lower bounds are almost the same even considering QES because the Earth-stopping effect is very weak for the small scattering cross section. These observations are also applicable to the vector mediator case, such as dark photon. The full Monte Carlo simulation may improve the Earth-stopping model but will not change our conclusions.

\section{Conclusion}

The inelastic DM-Earth scattering in the Earth stopping effect is usually neglected, however, which can be the dominant contribution for accelerated sub-GeV DM in the high kinetic energy region. As a proof of concept, in this work, we for the first time calculate the atmospheric DM-nucleus quasi-elastic and deep inelastic scattering in the Earth-stopping and derive new bounds on the DM interactions. We find that the mediator mass will affect the relative size of the elastic, quasi-elastic and deep inelastic scattering cross sections.
Including the contribution of the inelastic scattering in the Earth-stopping effect will change the resulting upper bound on the DM-nucleus scattering by about one order of magnitude in the Xenon1T direct detection.

\section{acknowledgments}
We are grateful to Artur M. Ankowski for a useful discussion. This work is supported by the National Natural Science Foundation of China (NNSFC) under grants No. 12275134, No. 12275232, and No. 11835005.

\bibliography{refs}

 \end{document}